\documentclass[twocolumn]{aastex63}

\usepackage{subfigure}
\usepackage{natbib}
\usepackage{amsmath}
\usepackage{soul}

\received{November 26, 2020}
\revised{December 20, 2020}
\accepted{January 5, 2021}
\submitjournal{\apj}

\newcommand{\modelRef}{\{$D$, $V_h$, $V_z$, $B$\} }
\newcommand{\modelB}[4]{\{#1, #2, #3, #4\}}
\newcommand{\rA}{$\mathrm{\AA}$}

\usepackage{soul}

\begin{document}

\title{The Effects of Three-dimensional Radiative Transfer \\on the
  Resonance Polarization of the Ca {\sc i} 4227 {\rA} line}

\shorttitle{The Effects of 3D Radiative Transfer on the 4227 {\rA}
  line polarization} \shortauthors{J. Jaume Bestard et al.}

\author[0000-0002-0970-5555]{J. Jaume Bestard}
\affiliation{Instituto de Astrof\'isica de Canarias, E-38205 La Laguna, Tenerife, Spain}
\affiliation{Departamento de Astrof\'isica, Universidad de La Laguna, E-38206 La Laguna, Tenerife, Spain}

\author[0000-0001-5131-4139]{J. Trujillo Bueno}
\affiliation{Instituto de Astrof\'isica de Canarias, E-38205 La Laguna, Tenerife, Spain}
\affiliation{Departamento de Astrof\'isica, Universidad de La Laguna, E-38206 La Laguna, Tenerife, Spain}
\affiliation{Consejo Superior de Investigaciones Cient\'ificas, Spain}

\author[0000-0002-8292-2636]{J. \v St\v ep\'an}
\affiliation{Astronomical Institute ASCR, Fric\v ova 298, 251 65 Ondr\v ejov, Czech Republic}

\author[0000-0003-1465-5692]{T. del Pino Alem\'an}
\affiliation{Instituto de Astrof\'isica de Canarias, E-38205 La Laguna, Tenerife, Spain}

\begin{abstract}
  The sizable linear polarization signals produced by the scattering
  of anisotropic radiation in the core of the Ca {\sc i} 4227 \r{A}\
  line constitute an important observable for probing the
  inhomogeneous and dynamic plasma of the lower solar chromosphere.
  Here we show the results of a three-dimensional (3D) radiative
  transfer complete frequency redistribution (CRD) investigation of
  the line's scattering polarization in a magneto-hydrodynamical 3D
  model of the solar atmosphere.  We take into account not only the
  Hanle effect produced by the model's magnetic field, but also the
  symmetry breaking caused by the horizontal inhomogeneities and
  macroscopic velocity gradients.  The spatial gradients of the
  horizontal components of the macroscopic velocities produce very
  significant forward scattering polarization signals without the need
  of magnetic fields, while the Hanle effect tends to depolarize them
  at the locations where the model's magnetic field is stronger than
  about 5 G.  The standard 1.5D approximation is found to be
  unsuitable for understanding the line's scattering polarization, but
  we introduce a novel improvement to this approximation that produces
  results in qualitative agreement with the full 3D results.  The
  instrumental degradation of the calculated polarization signals is
  also investigated, showing what can we expect to observe with the
  Visible Spectro-Polarimeter at the upcoming Daniel K. Inouye Solar
  Telescope.
\end{abstract}

\keywords{polarization -- radiative transfer -- scattering -- Sun: chromosphere -- Stars: chromospheres}

\section{Introduction} 

In order to probe the thermal and magnetic properties of the solar
chromospheric plasma we rely on the observation and interpretation of
the intensity and polarization spectra of strong resonance lines. Of
particular interest is the chromospheric line of Ca {\sc i} at 4227
{\rA} \cite[see][and references therein]{2018ApJ...854..150A}. This
resonance line shows the largest scattering polarization amplitude of
the solar visible spectrum \cite[e.g.,][]{2002sss..book.....G} and its
line-center signals are sensitive to the presence of magnetic fields
in the low solar chromosphere via the Hanle effect. Since the critical
magnetic field for the onset of the Hanle effect in the Ca {\sc i}
4227 \r{A}\ line is $B_{\rm H}=25$ G, the linear polarization at the
line-center is sensitive to magnetic fields with strengths between 5
and 125~G.

The new generation of solar telescopes will hopefully allow us to
measure, with unprecedented spatio-temporal resolution, the linear
polarization that the scattering of anisotropic radiation in the solar
atmosphere introduces in the emergent spectral line radiation. In
particular, with new telescopes like the 4-m Daniel K. Inouye Solar
Telescope \cite[DKIST;][]{rimmele_2020} it should be possible to
measure the scattering polarization of the Ca {\sc i} 4227 \r{A}\ line
with a spatial resolution of the order of 0\farcs1 and a temporal
resolution of the order of 10 seconds. The planning and interpretation
of such spectro-polarimetric observations requires a good theoretical
understanding of the physical mechanisms that produce polarization in
the Ca {\sc i} 4227 \r{A}\ line.

The radiative transfer investigations that have been carried out so
far used the one-dimensional (1D) approximation, either because they
are based on calculations in plane-parallel semi-empirical models of
the solar atmosphere
\cite[e.g.][]{1992A&A...258..521F,2011ApJ...737...95A,2014ApJ...793...42S,2018ApJ...854..150A}
or because the effects of horizontal radiative transfer were neglected
when using three-dimensional dynamical models
\cite[e.g.][]{2013ApJ...772...90L,2013ApJ...772...89L,2013ApJ...778..143P,2017ApJ...843...64C}. Such
theoretical investigations, based on the 1D radiative transfer
approximation, have been very important to understand that partial
frequency redistribution (PRD) produces very significant signals in
the wings of the scattering polarization $Q/I$ profiles
\cite[e.g.,][]{1973SoPh...28..271D, 1980A&A....88..302A,1982A&A...115....1R,1988A&A...194..268F}, 
that the
magneto-optical terms that couple Stokes $Q$ and $U$ produce an
interesting magnetic sensitivity in the wings of the $Q/I$ and $U/I$
profiles \cite[][]{2018ApJ...854..150A}, that the approximation of
complete frequency redistribution (CRD) is suitable to estimate the
line-center signals
\cite[e.g.,][]{2010ApJ...722.1269S,2011ApJ...737...95A}, and
that the spatial gradients of the model's vertical macroscopic
velocities can enhance and distort the scattering polarization
profiles \cite[][]{2012ApJ...751....5C,2013ApJ...764...40C,2015ApJ...812...28S,2019ApJ...879...48M}.

In 1D models of the solar atmosphere, either static or with only
vertical velocities, the radiation field illuminating each point
within the medium has axial symmetry around the local
vertical. Therefore, due to the ensuing axial symmetry, in the absence
of any inclined magnetic field there is no scattering polarization
when calculating the emergent spectral line radiation at the solar
disk-center (i.e., there is no forward scattering polarization). Under
such circumstances the only way to generate forward scattering
polarization is via the Hanle effect produced by an inclined magnetic
field
\cite[e.g.,][]{2001ASPC..236..161T,2002Natur.415..403T}. However, the
solar atmosphere is highly inhomogeneous and dynamic. The horizontal
inhomogeneities (e.g., in the plasma temperature and density) break
the axial symmetry of the incident radiation field; therefore, in
principle, we expect forward scattering polarization without the need
of an inclined magnetic field
\cite[e.g.,][]{2011ApJ...743...12M}. Moreover, the Doppler shifts
resulting from spatial gradients in the plasma macroscopic velocity
can also break the axial symmetry of the incident spectral line
radiation \cite[][]{2016ApJ...826L..10S,
  2018ApJ...863..164D}. Clearly, it is important to carefully study
the effect on the scattering line polarization signals of such
non-magnetic symmetry breaking.

The aim of this paper is to investigate this complex three-dimensional
(3D) radiative transfer problem for the resonance line of Ca {\sc i}
at 4227 \r{A}, focusing on the emergent spectral line radiation at the
solar disk-center (i.e., line of sight with $\mu=\cos{\theta}=1$,
$\theta$ being the heliocentric angle). The intensity and polarization
of this strong resonance line with angular momentum $J_l=0$ and
$J_u=1$ for the lower and upper levels, respectively, can be
approximately modeled using a two-level model atom, once the number
density of neutral calcium has been calculated by solving the standard
non-LTE radiative transfer problem for a realistic multilevel model
atom for Ca {\sc i} + Ca {\sc ii}. Our calculations assume CRD, which
is suitable for estimating the scattering polarization at the center
of the Ca {\sc i} 4227 \r{A}\ line where the Hanle effect
operates. Both, the two-level model and the CRD approximations make
feasible to carry out many numerical experiments applying the 3D
radiative transfer code PORTA\footnote{PORTA is a public code. The
  last version can be found at \url{https://gitlab.com/polmag/PORTA}}
of \cite{2013A&A...557A.143S}, which takes into account all the
above-mentioned symmetry breaking effects.

After formulating the problem in \S2, in \S3 we carry out a basic
radiative transfer investigation based on 1D models to clarify the
symmetry breaking effects produced by the spatial gradients of the
horizontal components of the macroscopic velocities.  The following
sections focus on a number of radiative transfer calculations in a 3D
snapshot model of the solar atmosphere resulting from a radiation
magneto-hydrodynamic simulation of an enhanced network region; such
carefully planned numerical experiments allow us to show and
understand the impact of the various competing symmetry breaking
effects. In particular, in \S4 we compare the full 3D results with
those obtained using the so-called 1.5D strategy, both considering the
three components of the velocity field and only the vertical
component. In \S5 we highlight the importance of the symmetry breaking
produced by the spatial gradients of the horizontal components of the
plasma macroscopic velocity, showing that it produces measurable
forward scattering polarization signals without the need of magnetic
fields. The impact of the Hanle effect caused by the model's magnetic
field is discussed in \S6, showing that it tends to produce
depolarization at the most magnetized locations. In \S7 we study the
instrumental degradation of the theoretical polarization signals,
illustrating what we may observe with new instruments like the Visible
Spectro-Polarimeter (ViSP) attached to the DKIST. Finally, a summary
with our conclusions can be found in \S8.

\section{Formulation of the problem\label{sec:formulation}}

\begin{figure*}[]
  \centering
  \epsscale{1.05}
  \plotone{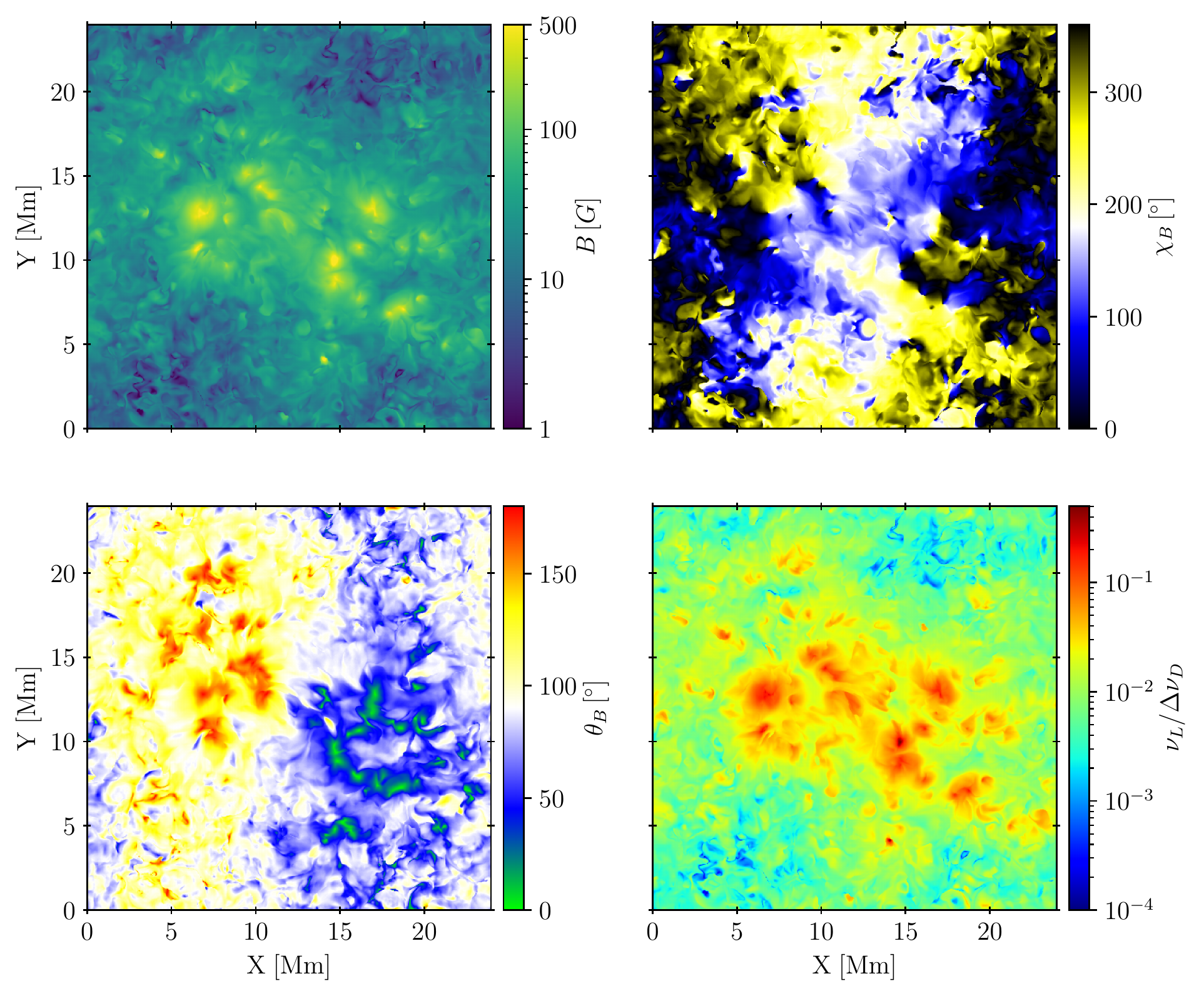}
  \caption{{\bf Top left}: magnetic field strength. {\bf Top right}:
    magnetic field azimuth. {\bf Bottom left}: magnetic field
    inclination. {\bf Bottom right}: ratio of the Larmor frequency
    over the thermal width of the line. All quantities are taken at
    the corrugated surface within the 3D model where the optical depth
    at the Ca {\sc i} 4227 \r{A}\ line-center is unity along the
    vertical line of sight.}
  \label{fig:Btau}
\end{figure*}

The linear polarization of the Ca {\sc i} 4227 \r{A}\ line-center
outside active regions is dominated by scattering processes and the
Hanle effect. In this paper, we focus on theoretically investigating
such spectral line polarization taking into account the effects of 3D
radiative transfer. To this end, we consider a 3D model of an enhanced
network region resulting from a radiation magneto-hydrodynamic
simulation \cite[see][]{2016A&A...585A...4C} and calculate the Stokes
profiles of the emergent radiation by applying the 3D radiative
transfer code PORTA \citep{2013A&A...557A.143S}.

PORTA solves the non-LTE problem of the generation and transfer of
spectral line polarization neglecting frequency correlations
between the incoming and outgoing photons in the scattering events
(see the relevant equations in section 7.2 of
\citealt{2004ASSL..307.....L}). This complete frequency redistribution
limit is suitable for estimating the scattering polarization at the
line-center \cite[e.g.,][]{2010ApJ...722.1269S} where the Hanle effect
operates.

The upper-left panel in {\figurename} \ref{fig:Btau} shows the
magnetic field strength of the 3D snapshot model atmosphere at the
heights where the line-center optical depth along the $\mu=1$ line of
sight (LOS) is unity.  The bottom left panel, showing the magnetic
field inclination with respect to the local vertical, indicates that
the 3D model presents two patches of opposite magnetic polarity
separated by about 8 Mm, where the magnetic field strength reaches
values larger than 200 G (see the upper left panel).  The model's
magnetic field lines joining such two patches have a dominant azimuth
(see the blue color in the upper right panel) and they reach
chromospheric and coronal heights. Even at the relatively strong field
locations of such patches the Zeeman splitting of the line's upper
level, as quantified by the Larmor frequency $\nu_L$, is smaller than
the line's Doppler width (see the lower right panel).

The dimensions of the 3D snapshot model atmosphere are
$24\times24\times16.8\, \rm{Mm}^3$, going from $2.4\, \rm{Mm}$ below
the average height of continuum optical depth unity to
$14.4\, \rm{Mm}$ above.  We point out that the atmospheric heights
where the line-center optical depth of the Ca {\sc i} 4227 \r{A}\ line
is unity lie several hundreds of kilometers below the model's
chromosphere-corona transition region. Accordingly, for the solution
of the corresponding radiative transfer problem it was more than
sufficient to restrict the vertical extension of the model from
$0.1\, \rm{Mm}$ below the model's visible surface to $2.5\, \rm{Mm}$
above.

Once the number density of Ca {\sc i} is known at each spatial grid
point, the Stokes profiles of the Ca {\sc i} 4227 \r{A}\ resonance
line are calculated with PORTA using the two-level model atom
approximation.  This approximation is suitable because the level
$^1{\rm P}_1^{\circ}$ of the Ca {\sc i} atom, the upper level of the
4227 {\rA} line, is not radiatively coupled to any other lower level
besides the ground level $^1{\rm S}_0$, the line's lower level.  The
Einstein coefficient for spontaneous emission from the line's upper
level (angular momentum $J_u=1$ and Land\'e factor $g_u=1$) to the
line's lower level (angular momentum $J_{\ell}=0$) is
$A_{ul}=2.18{\times}10^8\,\rm{s^{-1}}$. The rate of inelastic
collisions with electrons from the upper level to the lower level
($C_{u\ell}$) was calculated following \cite{1962Obs....82..111S} and
the damping parameter was computed including the collisional
broadening and the quadratic Stark effect
\cite[][]{1998PASA...15..336B}. All these parameters, the background
continuum quantities (opacity and emissivity) and the number density
of Ca {\sc i} atoms at each spatial grid point were computed with the
RH code \cite[][]{2001ApJ...557..389U} using a more realistic atomic
model with 20 Ca {\sc i} levels and the ground level of Ca {\sc ii}.

The critical magnetic field $B_H$ for the onset of the Hanle effect is
25 G ($B_H\approx1.137\times10^{-7}/(t_{\rm{life}}\,g_u)$;
{$t_{\mathrm{life}}$ and $g_u$ being the lifetime of the upper level 
in seconds and its Land\'e factor, respectively}). Because
collisional depolarization is not significant for the core of this
line, there is no collisional quenching
\cite[e.g.,][]{2018ApJ...854..150A}. Since the lower level has
$J_{\ell}=0$, the scattering polarization in this line is solely due
to the radiatively-induced atomic polarization of the upper level
(i.e., to the population imbalances and quantum coherence between the
magnetic sublevels of the line's upper level with $J_u=1$).

\section{Symmetry breaking by horizontal macroscopic velocities\label{sec:SB}}

\begin{figure*}[t]
  \centering
  \epsscale{1.18}
  \plotone{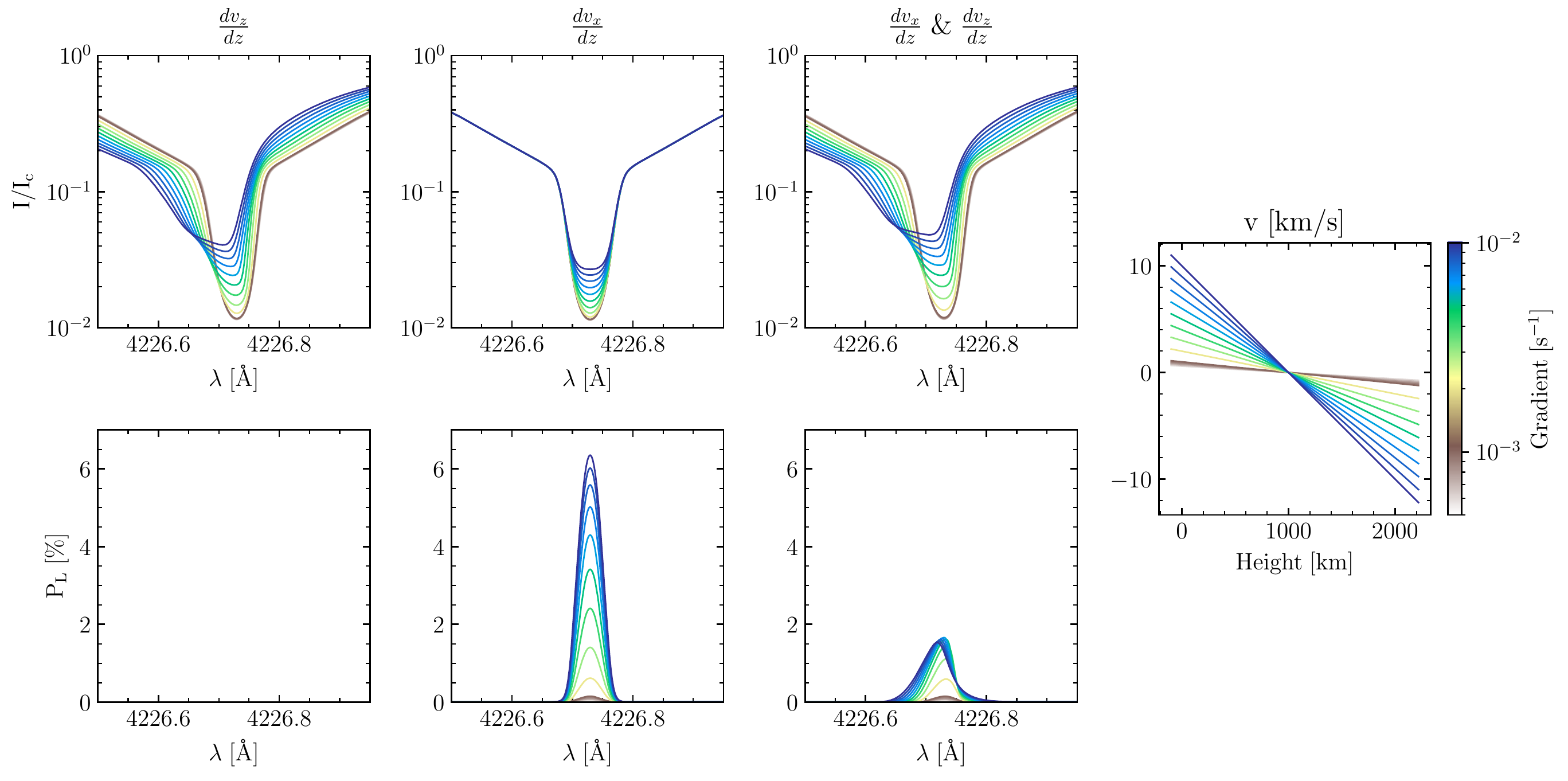}
  \caption{Emergent intensity (top row) and $P_L$ (bottom row)
      in the Ca {\sc i} 4227 {\rA} line calculated in the FAL-C
      semi-empirical model for the disk-center LOS. \textbf{First
        column:} considering vertical gradients in the vertical
      velocity. \textbf{Second column:} vertical gradients in the
      horizontal velocity component (along the X axis). \textbf{Third
        column:} vertical gradients in both components. Each velocity
      component is defined by a linear $v_{x,z}(z)$ function
      decreasing with height with slopes between $5\cdot10^{-4}$ and
      $10^{-2}~\rm{s^{-1}}$, fulfilling $v_{x,z}(1000) = 0$ (right
      panel). Positive vertical velocities indicates that the plasma
      is moving towards the observer.}
  \label{fig:gradients}
\end{figure*}

When the polarization of the spectral line radiation is ignored and
one considers only the Stokes $I$ parameter, the relevant statistical
equilibrium equations are those for the overall atomic level
populations \cite[e.g., ][]{1978stat.book.....M} and the radiation
field is fully characterized by the mean intensity:
\begin{equation}
    \bar{J^0_0} = \int d{\nu}\oint \frac{d\vec{\Omega}}{4\pi} {\phi\left(\nu\left[1-\frac{\vec{v}\cdot\vec{\Omega}}{c}\right]\right)}
    \,I_{\nu \vec{\Omega}}~,
    \label{E-J00}   
  \end{equation}
where $\phi$ is the Voigt absorption profile, $\nu$ the frequency,
$\vec{v}$ the macroscopic plasma velocity and $\vec{\Omega}$ the
propagation direction of the radiation beam. This direction is
characterized by $\mu=\cos\theta$ (with $\theta$ the inclination of
the ray with respect to the local vertical) and the azimuth $\chi$.

When scattering polarization is accounted for, the relevant
statistical equilibrium equations are those for the density matrix
elements $\rho^K_Q(J)$ for each atomic level with angular momentum $J$
\cite[see][section 7.2]{2004ASSL..307.....L}. The mean intensity is no
longer enough to characterize the radiation field, requiring the
definition of the radiation field tensor $\bar{J}^K_Q$ with
components:

\begin{widetext}
\begin{subequations}\label{E-JKQ}
\begin{align}
  \bar{J}^2_0&= \frac{1}{2\sqrt{2}} \int d\nu\oint \frac{d \vec{\Omega}}{4\pi} {\phi\left(\nu\left[1-\frac{\vec{v}\cdot\vec{\Omega}}{c}\right]\right)}
  \Big[(3\mu^2-1)I_{\nu \vec{\Omega}}+3(\mu^2-1)Q_{\nu
  \vec{\Omega}}\Big],
                                                            \label{E-J20}
\displaybreak[0]\\
  {\rm Re}[{\bar J}^2_1] &= \frac{\sqrt{3}}{2} \int d\nu \oint \frac{d\vec{\Omega}}{4\pi} {\phi\left(\nu\left[1-\frac{\vec{v}\cdot\vec{\Omega}}{c}\right]\right)}
  \sqrt{1-\mu^2}
  \Big[-\mu\cos\chi(I_{\nu \vec{\Omega}}+Q_{\nu\vec{\Omega}})+\sin\chi U_{\nu\vec{\Omega}}\Big],
                                                            \label{E-RJ21}
\displaybreak[0]\\
  {\rm Im}[{\bar J}^2_1] &= \frac{\sqrt{3}}{2} \int d\nu \oint \frac{d\vec{\Omega}}{4\pi} {\phi\left(\nu\left[1-\frac{\vec{v}\cdot\vec{\Omega}}{c}\right]\right)}
  \sqrt{1-\mu^2}\Big[-\mu\sin\chi(I_{\nu \vec{\Omega}}+Q_{\nu
  \vec{\Omega}})-\cos\chi U_{\nu \vec{\Omega}}\Big],
                                                            \label{E-IJ21}
\displaybreak[0]\\
  {\rm Re}[{\bar J}^2_2] &= \frac{\sqrt{3}}{4}\int d\nu \oint \frac{d\vec{\Omega}}{4\pi}\, {\phi\left(\nu\left[1-\frac{\vec{v}\cdot\vec{\Omega}}{c}\right]\right)}
  \bigg[\cos(2\chi)\Big[(1-\mu^2)I_{\nu \vec{\Omega}}-(1+\mu^2)Q_{\nu
  \vec{\Omega}}\Big]+2\sin(2\chi)\mu U_{\nu \vec{\Omega}}\bigg],
                                                            \label{E-RJ22}
\displaybreak[0]\\
  {\rm Im}[{\bar J}^2_2] &= \frac{\sqrt{3}}{4}\int d\nu \oint \frac{d\vec{\Omega}}{4\pi}\, {\phi\left(\nu\left[1-\frac{\vec{v}\cdot\vec{\Omega}}{c}\right]\right)}
  \bigg[\sin(2\chi)\Big[(1-\mu^2)I_{\nu \vec{\Omega}}-(1+\mu^2)Q_{\nu
  \vec{\Omega}}\Big]-2\cos(2\chi)\mu U_{\nu \vec{\Omega}}\bigg],
                                                            \label{E-IJ22}
\end{align}
\end{subequations}
\end{widetext}
where $I_{\nu\vec{\Omega}}$, $Q_{\nu\vec{\Omega}}$, and
$U_{\nu\vec{\Omega}}$ are the Stokes parameters for the given
frequency ($\nu$) and direction ($\vec{\Omega}$).  In these
expressions the reference direction for positive $Q_{\nu\vec{\Omega}}$
is in the plane formed by $\vec{\Omega}$ and the local vertical.  The
$\bar{J}^2_0$ component quantifies the anisotropy of the incident
radiation. The real and imaginary parts of the ${\bar J}^2_1$ and
${\bar J}^2_2$ components quantify the breaking of the axial symmetry
of the incident radiation field with respect to the local vertical.

It is known that the spatial gradients of the vertical component of
the macroscopic velocity can modify the values of the $\bar{J}^2_0$
anisotropy tensor and, therefore, the amplitude and shape of the
scattering polarization profiles
\cite[see][]{2012ApJ...751....5C,2013ApJ...764...40C}.  Even more
important is to note that the spatial gradients in the {\em
  horizontal} components of the macroscopic velocity break the axial
symmetry of the incident radiation field
\cite[][]{2016ApJ...826L..10S, 2018ApJ...863..164D}, even in an
unmagnetized 1D model atmosphere. The ensuing non-zero values for the
${\bar J}^2_1$ and ${\bar J}^2_2$ tensors dramatically impact the
$Q/I$ and $U/I$ signals, giving rise to forward scattering
polarization without the need of inclined magnetic fields. In order to
illustrate this, in {\figurename} \ref{fig:gradients} we show the
intensity and the total fractional linear polarization
($P_L=\sqrt{Q^2+U^2}/I$) profiles of the Ca {\sc i} 4227 \r{A}\ line
for the disk-center LOS calculated in the semi-empirical model C of
\citet{1993ApJ...406..319F}, which is unmagnetized (hereafter, FAL-C
model). We included macroscopic velocity fields with constant
gradients in the vertical and horizontal components. Note that in a
plane-parallel model atmosphere the only possible gradients are along
the vertical direction. The right panel of {\figurename}
\ref{fig:gradients} shows the different macroscopic velocity gradients
we have considered in the FAL-C model. Such academic velocity fields
have zero velocity at the height where the optical depth is unity at
the center of the Ca {\sc i} 4227 \r{A}\ line.  We have opted for this
curious choice to more easily emphasize the impact of the spatial
gradients in the macroscopic velocity. When we only allow for the
vertical component of the macroscopic velocity (first column), the
intensity profiles are deformed due to different Doppler shifts at
different atmospheric heights, and because the gradient is negative
such deformation is more prominent in the blue part of the Stokes-$I$
profile. Clearly, the linear polarization at the solar disk center is
zero, because the Doppler shifts caused by the vertical velocities are
axially symmetric. However, there is a significant amount of
forward-scattering polarization when we only consider the horizontal
velocity component (second column).  As expected, due to the
azimuthally non-symmetric Doppler shifts, the stronger the gradient
the larger the polarization signal. The line-center intensity becomes
larger and flatter because of the broadening produced by the gradients
in the horizontal velocity. When the gradients in both velocity
components are included, the signal is much smaller and the shapes of
the profiles are modified due to the impact of the Doppler shifts on
the intensity profiles (third column).
  
\section{An improved 1.5D approximation\label{sec:1.5D}}

The so-called 1.5D approximation for simplifying the numerical
solution of non-LTE radiative transfer problems in 3D models of
stellar atmospheres consists in neglecting the impact of horizontal
radiative transfer on the excitation state of the atomic system. In
other words, for each vertical column of the 3D model under
consideration one solves the non-LTE radiative transfer problem as if
such column was an independent 1D plane-parallel atmosphere.

To the best of our knowledge, the 1.5D approximation has always been
applied neglecting the horizontal components of the model's
macroscopic velocity at each iterative step in the non-LTE radiative
transfer calculation \cite[e.g.,][]{2013ApJ...772...90L,
  2013ApJ...772...89L, 2013ApJ...778..143P, 2015ApJ...801...16C,
  2017ApJ...843...64C}. Under such circumstances, only the presence of
magnetic fields inclined with respect to the local vertical can break
the axial symmetry of the incident radiation field at each point
within the model atmosphere. Consequently, only the Hanle effect due
to the inclined magnetic fields can produce $Q/I$ and $U/I$ forward
scattering (disk center) polarization signals. In other words, under
such assumptions, the detection of scattering polarization at the
solar disk-center implies the presence of an inclined magnetic
field. While this is a valid conclusion when the radiation field that
pumps the atomic system has axial symmetry around the local vertical
(e.g., quiescent coronal filaments observed in the He {\sc i} 10830
{\rA} multiplet, see \citealt{2002Natur.415..403T}), it is not for
spectral lines that originate within the inhomogeneous and dynamic
plasma of the solar atmosphere itself.

The 1.5D approximation can never take into account the breaking of the
axial symmetry of the pumping radiation field that results from the
horizontal thermal and density inhomogeneities in the 3D
model. Although this can only be accounted for via full 3D radiative
transfer calculations, such as those reported below, the spatial
gradients of the horizontal velocity components are an additional
source of non-magnetic symmetry breaking that can be considered in the
1.5D approximation (the Doppler shift of these velocity components
modify the symmetry properties of the incident radiation field; e.g.,
{\figurename} 11 in \citealt{2018ApJ...863..164D}, which shows the
forward scattering polarization in the Sr {\sc i} 4607 {\rA} line). By
taking into account the horizontal components of the macroscopic
velocity at each height for each individual column of the 3D model
atmosphere we obtain forward scattering signals without the need of a
magnetic field; i.e., we are able to account for the symmetry breaking
that results from the vertical gradients of the horizontal component
of the model's macroscopic velocity.

To facilitate the identification of each particular radiative transfer
solution we use the notation \modelRef where, for example,
\modelB{1}{1}{1}{1} indicates the most realistic case of a full 3D
solution ($D=1$) that takes into account the impact of the horizontal
($V_h=1$) and vertical ($V_z=1$) macroscopic velocities, and the Hanle
effect produced by the model's magnetic field ($B=1$), while
\modelB{0}{0}{0}{0} refers to the 1.5D solution ($D=0$) ignoring the
model's horizontal ($V_h=0$) and vertical ($V_z=0$) macroscopic
velocities and its magnetic field ($B=0$).

\begin{figure*}[ht!]
  \centering
  \epsscale{1.15}
  \plotone{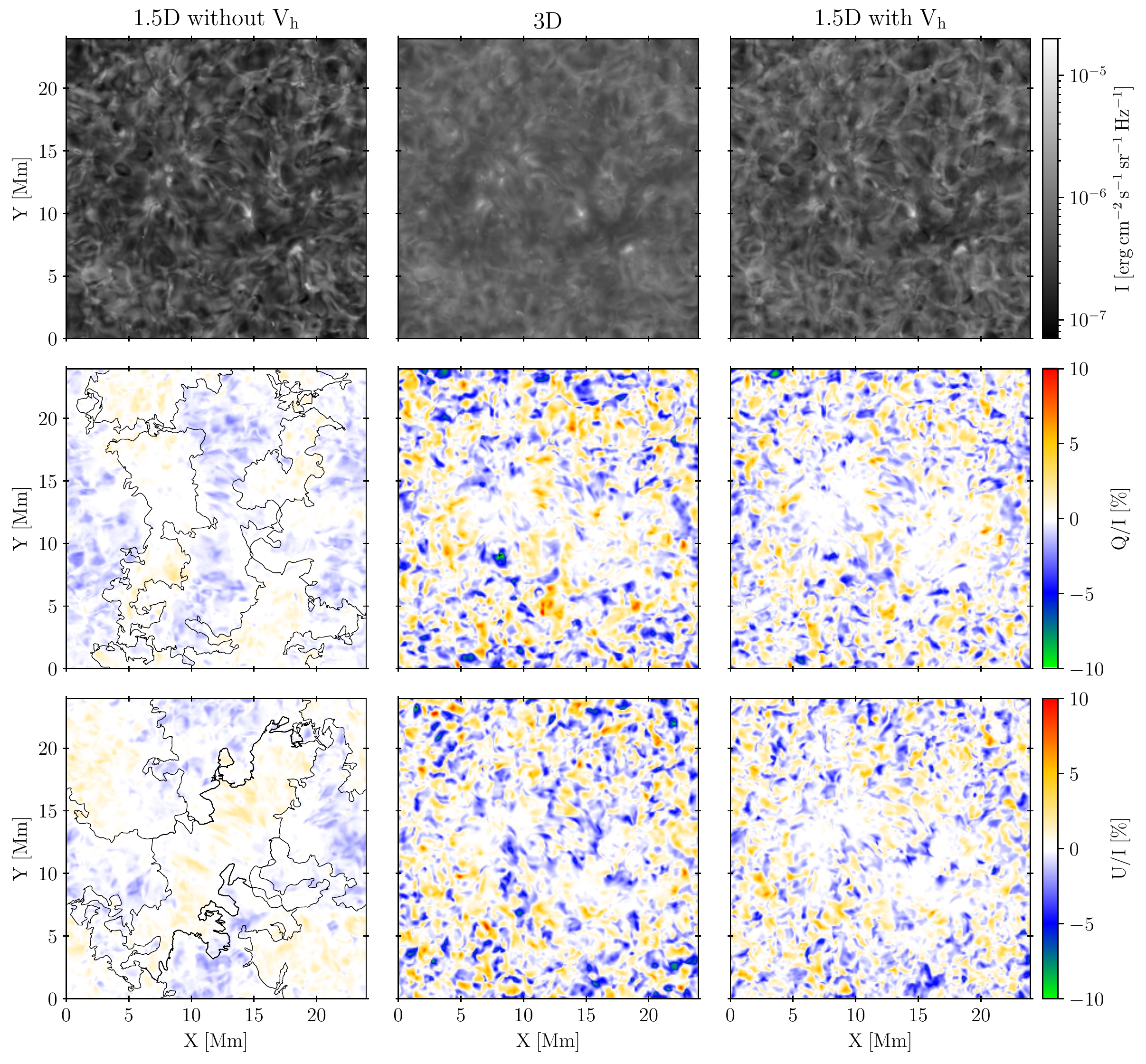}
  \caption{Values of the emergent Stokes $I$ (top row), $Q/I$ (middle
    row) and $U/I$ (bottom row) disk-center signals for the
    \modelB{0}{0}{1}{1} (left column), \modelB{1}{1}{1}{1} (middle
    column), and \modelB{0}{1}{1}{1} (right column) cases (see main
    text). The black contours show the positions where the model's
    magnetic field azimuth has any of the following specific values at
    the heights where $\tau(\lambda_{P_L})=1$ (we recall that
    $\lambda_{P_L}$ is the wavelength at which $P_L$ has its maximum
    value).  In the $Q/I$ panel such specific values are
    $\chi_B=45^{\circ}$, $135^{\circ}$, $225^{\circ}$, and
    $315^{\circ}$, while in the $U/I$ panel they are
    $\chi_B=0^{\circ}$, $90^{\circ}$, $180^{\circ}$, and
    $270^{\circ}$. We point out that for improving the visualization
    we have retained only the large-scale contours.  The positive
    reference direction for Stokes $Q$ is along the Y-axis of the
    figures. The scale is saturated at $7\cdot10^{-7}$ and
    $2\cdot10^{-5}$ cgs for the intensity and at $\pm10\%$ for $Q/I$
    and $U/I$.}
  \label{fig:IQU_1.5D0Vh_1.5D_3D}
\end{figure*}

For each point of the field of view (FOV), {\figurename}
\ref{fig:IQU_1.5D0Vh_1.5D_3D} shows the emergent Stokes $I$, $Q/I$,
and $U/I$ disk-center signals at the wavelength $\lambda_{P_L}$ where
the total linear polarization $P_L$ is maximum (hereafter, the $I$,
$Q/I$, $U/I$ and $P_L$ signals). The various panels of the figure show
such signals for (a) the 1.5D approximation neglecting the horizontal
components of the macroscopic velocity (left column), (b) the full 3D
solution (middle column), and (c) the 1.5D approximation taking into
account the horizontal components of the velocity (right column). The
self-consistent solutions have been obtained applying PORTA till
reaching a maximum relative change smaller than $10^{-4}$ in the
$\rho^0_0(J)$ density matrix elements. The emergent Stokes profiles
have been calculated taking into account the Doppler shifts produced
by the model's macroscopic velocities along the LOS. Therefore,
discrepancies in the calculated fractional polarization signals are
solely due to the differences in the iteratively computed density
matrix components $\rho^K_Q(J)$.
             
The left column of {\figurename} \ref{fig:IQU_1.5D0Vh_1.5D_3D} shows
the $I$, $Q/I$, and $U/I$ signals for \modelB{0}{0}{1}{1}. The
resulting $Q/I$ and $U/I$ disk-center signals are fully due to the
Hanle effect caused by the model's magnetic field, as there is no
forward scattering polarization in the non-magnetic 1.5D case when the
horizontal components of the macroscopic velocities are neglected (see
the second panel of {\figurename} \ref{fig:gradients}). We see a $Q/I$
($U/I$) pattern where the sign of $Q/I$ ($U/I$) is different inside
and outside the contours indicated in the $Q/I$ ($U/I$) left panel of
the figure. These contours in this $Q/I$ ($U/I$) panel indicate the
points of the FOV where the model's magnetic field azimuth (see
{\figurename} \ref{fig:Btau}) is such that ${\rm cos}{2\chi_B}=0$
(${\rm sin}{2\chi_B}=0$), which is a factor appearing in an
approximate equation for $Q/I$ ($U/I$) valid only for spectral lines
in the saturation regime of the Hanle effect and whenever an inclined
magnetic field is the only possible cause of symmetry breaking
\citep[e.g.,][]{Casini-Landi-review,2010ASSP...19..118T,
  2015ApJ...801...16C}.

The middle column of {\figurename} \ref{fig:IQU_1.5D0Vh_1.5D_3D} shows
the emergent Stokes $I$, $Q/I$, and $U/I$ signals for the
\modelB{1}{1}{1}{1} 3D solution, which accounts for all possible
causes of symmetry breaking (magnetic and non-magnetic). In addition
to the smoothing effect of horizontal radiative transfer, clearly seen
in the intensity (top panel), the $Q/I$ and $U/I$ polarization
patterns show more structure, with positive and negative signals
covering relatively small patches all over the FOV, in contrast with
the larger scale pattern of the \modelB{0}{0}{1}{1} solution (left
column).

The right column of {\figurename} \ref{fig:IQU_1.5D0Vh_1.5D_3D} shows
the emergent Stokes $I$, $Q/I$, and $U/I$ signals for the improved
\modelB{0}{1}{1}{1} 1.5D solution, which includes the symmetry
breaking caused by the vertical gradients of the horizontal components
of the macroscopic plasma velocity. Interestingly, despite being a
1.5D solution, accounting for such non-magnetic symmetry breaking
effects is sufficient to produce forward scattering polarization
patterns qualitatively similar to those corresponding to the full
\modelB{1}{1}{1}{1} 3D solution. However, there are some significant
quantitative differences; e.g., while the spatially-averaged total
linear polarization signal is $\langle P_L \rangle=2.3~\%$ in
\modelB{1}{1}{1}{1}, it is $\langle P_L \rangle=1.8~\%$ in
\modelB{0}{1}{1}{1}. This difference is due to the fact that only the
full 3D solution can account for the symmetry breaking produced by the
model's horizontal inhomogeneities.

\begin{figure*}[]
  \centering
  \epsscale{1.15}
  \plotone{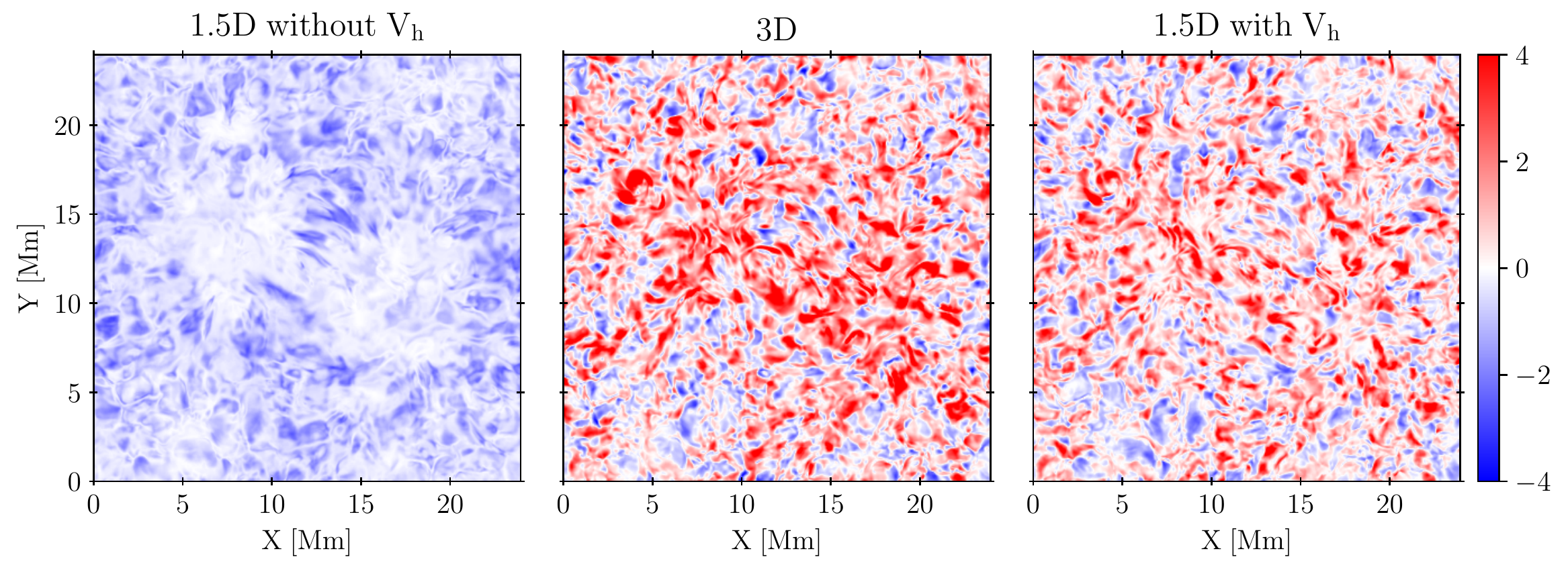}
  \caption{Pixel by pixel differences between the 
    total linear polarization $P_L$ signal calculated without and with the
    model's magnetic field, $P_L(B=0)-P_L$. {\bf Left panel}: 1.5D
    approximation without $V_h$. {\bf Middle panel}: full 3D
    calculation. {\bf Right panel}: 1.5D approximation with $V_h$. The
    scale is saturated at $\pm4$\%.}
  \label{fig:PL_1.5D_B}
\end{figure*}

{\figurename} \ref{fig:PL_1.5D_B} quantifies the modification of the
total linear polarization signals due to the Hanle effect --that
is, it shows the differences in the $P_L$ signals when ignoring and
taking into account the model's magnetic field. The left panel shows
the 1.5D approximation without horizontal velocities
(\modelB{0}{0}{1}{0} vs \modelB{0}{0}{1}{1}), the middle panel the
full 3D calculation (\modelB{1}{1}{1}{0} vs \modelB{1}{1}{1}{1}), and
the right panel the 1.5D approximation with horizontal velocities
(\modelB{0}{1}{1}{0} vs \modelB{0}{1}{1}{1}). As expected, the
standard 1.5D approximation can only lead to the (generally wrong)
conclusion that the Hanle effect produces forward scattering
polarization (negative signs in the figure). However, the full 3D
calculation shows that the Hanle effect tends to depolarize its
non-magnetic forward scattering signals in the region where the
model's magnetic loops are concentrated (namely, the top-left
bottom-right diagonal). A similar conclusion can be extracted from the
right panel corresponding to the 1.5D calculation that includes the
horizontal components of the velocity, although the Hanle
depolarization is not as significant as in the correct 3D
solution. Outside the region of the 3D model where the magnetic field
is the strongest, it is equally likely to find pixels where the Hanle
effect enhances or reduces the forward scattering polarization, in
both the 3D and 1.5D solutions with horizontal velocities
included. The impact of the Hanle effect on the linear scattering
polarization depends on the spectral line under consideration
\citep[e.g., the review by][]{2015IAUS..305..360S}.

\section{The impact of macroscopic velocity gradients on the
  scattering line polarization\label{sec:V}}

\begin{figure*}
  \centering
  \epsscale{1.17}
  \plotone{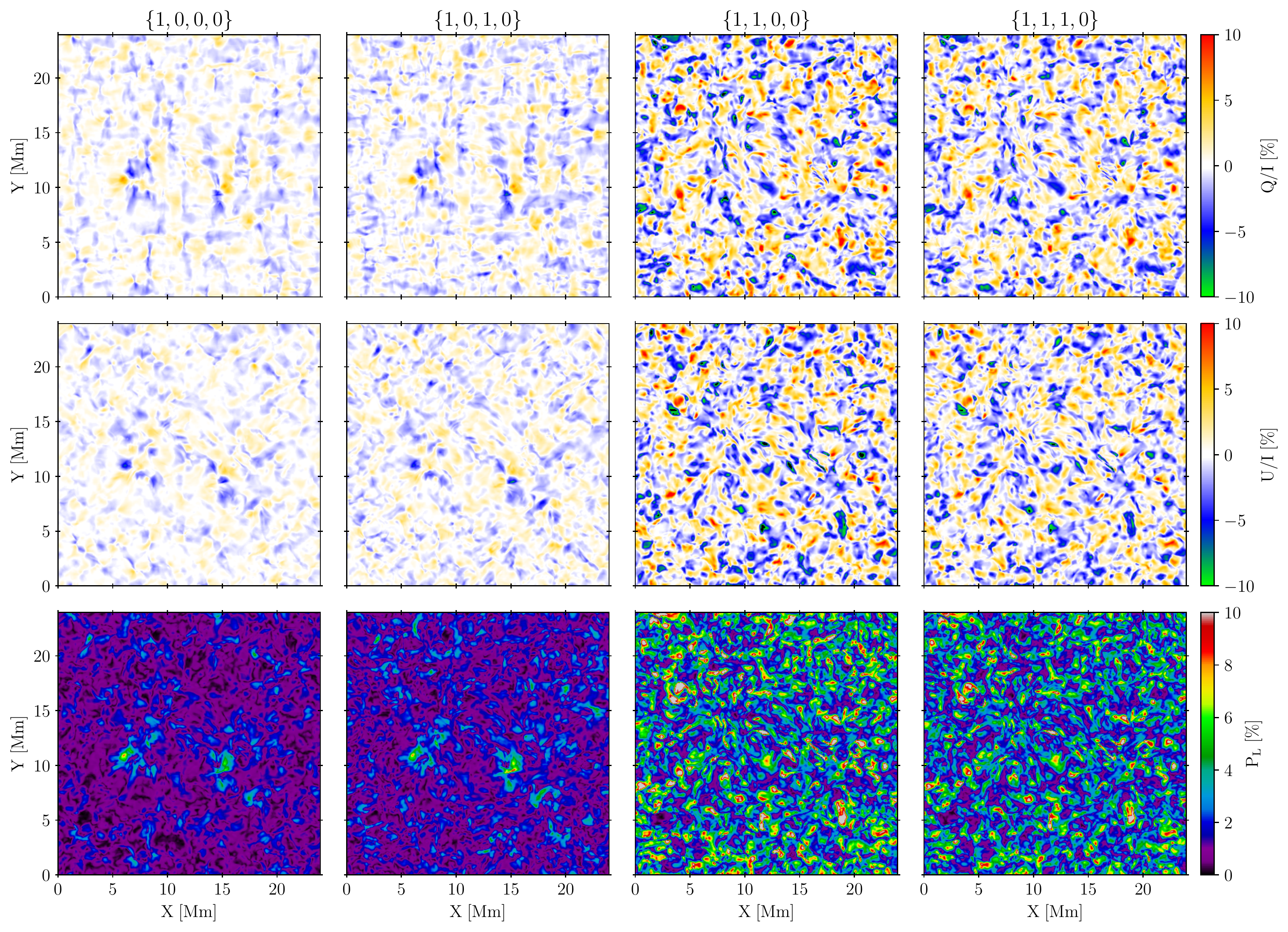}
  \caption{Non-magnetic $Q/I$ (top row), $U/I$ (middle row), and $P_L$
    (bottom row) disk-center signals of the Ca {\sc i} 4227 {\rA}
    line calculated by solving the 3D radiative transfer problem for
    the following cases: \modelB{1}{0}{0}{0} (first column),
    \modelB{1}{0}{1}{0} (second column), \modelB{1}{1}{0}{0} (third
    column), and \modelB{1}{1}{1}{0} (fourth column). The scales are
    saturated at $\pm 10~\%$. The positive reference direction for
    Stokes $Q$ is along the Y-axis of the figures.}
  \label{fig:QUP_0B0V_0B0Vh_0B0Vz}
\end{figure*}
\begin{figure*}
  \centering
  \epsscale{1}
  \plotone{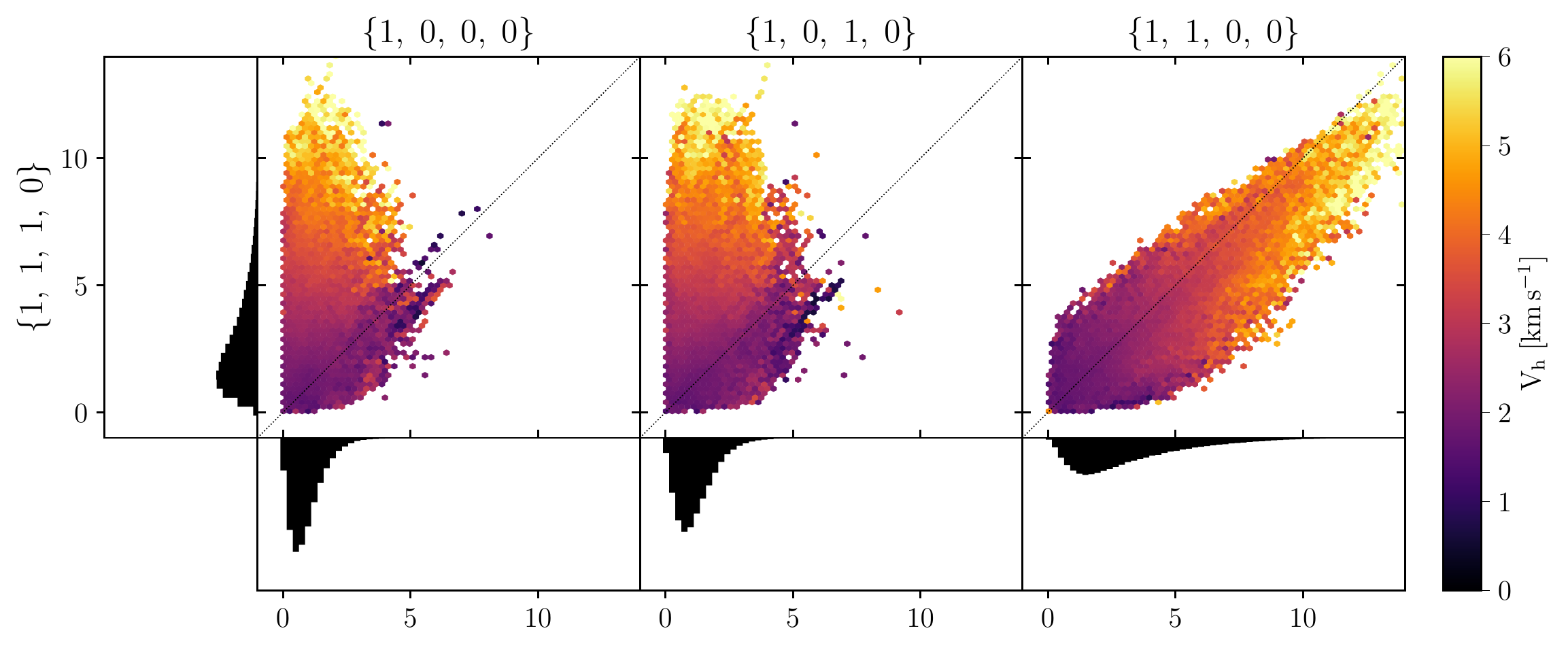}
  \caption{The $P_L$ signals (given in \%) of the
    \modelB{1}{1}{1}{0} case versus those corresponding to the
    \modelB{1}{0}{0}{0} (left panel), \modelB{1}{0}{1}{0} (middle
    panel), and \modelB{1}{1}{0}{0} (right panel) solutions. The color
    scale indicates the modulus of the horizontal component of the
    model's macroscopic velocities $V_h$ at the atmospheric heights
    where $\tau(\lambda_{P_L})=1$ in \modelB{1}{1}{1}{0}. In bins with
    more than one point the averaged $V_h$ is used. The diagonal
    dotted lines indicate where the points should be located in the
    case of a perfect correlation between the cases being
    compared. The black-colored areas at the left and at the bottom of
    the figure are 1D normalized histograms showing the distribution
    of $P_L$ values for each of the considered 3D non-magnetic
    solutions.}
  \label{fig:histV}
\end{figure*}

By means of full 3D radiative transfer calculations, in this section
we study the impact of the gradients in the vertical and horizontal
components of the model's macroscopic velocities on the forward
scattering polarization signals. Since we take into account the
effects of horizontal radiative transfer, we are automatically
accounting for the breaking of the axial symmetry of the pumping
radiation field, with respect to the local vertical, produced by the
horizontal thermal and density inhomogeneities in the 3D model
atmosphere. In the 3D numerical experiments shown in this section we
neglect the Hanle effect in order to isolate the non-magnetic sources
of symmetry breaking. Every emergent Stokes signal shown in this
section has been computed taking into account the Doppler shifts along
the LOS and, therefore, any difference between the various cases is
purely due to the impact of the velocity gradients on the density
matrix elements $\rho^K_Q(J)$, and not due to LOS Doppler shifts.

To study the impact of the spatial gradients in the vertical component
of the model's macroscopic velocities we first compare the
\modelB{1}{0}{0}{0} and \modelB{1}{0}{1}{0} cases. The axial symmetry
breaking in the former is due solely to the model's horizontal thermal
and density inhomogeneities. In the latter, the gradients of the
vertical component of the macroscopic velocity modify the
\modelB{1}{0}{0}{0} forward scattering signals. Secondly, we consider
the \modelB{1}{1}{0}{0} solution in order to reveal the dramatic
impact caused by the spatial gradients of the horizontal component of
the macroscopic velocities on the scattering polarization of the
emergent spectral line radiation. Finally, we consider the general
non-magnetic case (i.e., \modelB{1}{1}{1}{0}).

{\figurename} \ref{fig:QUP_0B0V_0B0Vh_0B0Vz} shows the $Q/I$, $U/I$,
and $P_L$ signals for the above-mentioned cases, from left to right:
\modelB{1}{0}{0}{0}, \modelB{1}{0}{1}{0}, \modelB{1}{1}{0}{0}, and
\modelB{1}{1}{1}{0}. The first and second columns show very similar
polarization patterns, thus indicating that the gradients of the
vertical component of the macroscopic velocities slightly modify the
forward scattering polarization signals
($\langle P_L \rangle \approx 1.0~\%$ in \modelB{1}{0}{0}{0} and
$\langle P_L \rangle \approx 1.3~\%$ in \modelB{1}{0}{1}{0}). Both,
the pattern and values of the polarization signals dramatically change
when the gradients of the horizontal component of the macroscopic
velocity are accounted for (\modelB{1}{1}{0}{0}, third column). In
this case, the whole FOV is covered by small-scale sign-changing $Q/I$
and $U/I$ forward scattering signals, with
$\langle P_L \rangle \approx 3.4~\%$. The full \modelB{1}{1}{1}{0}
non-magnetic 3D solution (forth column) is qualitatively similar to
the \modelB{1}{1}{0}{0} case (third column). The combined action of
the vertical and horizontal velocity components gives a smaller
average polarization, $\langle P_L \rangle \approx 3.0~\%$, compared
with the \modelB{1}{1}{0}{0} case.

In order to relate the forward scattering signals with the plasma
properties of the 3D model, we show in {\figurename} \ref{fig:histV}
scatter plots comparing pixel-by-pixel the $P_L$ signals of the
\modelB{1}{1}{1}{0} solution with the other three cases in
{\figurename} \ref{fig:QUP_0B0V_0B0Vh_0B0Vz}. The color table
indicates the absolute value of the macroscopic horizontal velocities,
$V_h=(v_x^2+v_y^2)^{1/2}$, at $\tau_{\mu=1}=1$ and
$\lambda_{P_L}$. Obviously, if the agreement (correlation) between two
solutions were perfect, the points in the scatter plot would be
located on the diagonal. From the left and middle panels, we see that
the \modelB{1}{0}{0}{0} and \modelB{1}{0}{1}{0} clearly underestimate
both the $P_L$ signals and range of variability ($P_L\lesssim10\%$
in \modelB{1}{1}{1}{0} while $P_L\lesssim5\%$ in \modelB{1}{0}{0}{0}
and \modelB{1}{0}{1}{0}). The right panel, comparing
\modelB{1}{1}{1}{0} and \modelB{1}{1}{0}{0}, shows a significant
degree of correlation. This correlation means that we obtain similar
$P_L$ signals from both solutions in the same pixels of the
FOV. Interestingly, the vertical components of the macroscopic
velocities slightly reduce the forward scattering signals
generated by the axial symmetry breaking caused by the horizontal
components of the macroscopic velocities. Furthermore, a careful attention to the
color table of {\figurename} \ref{fig:histV} reveals that there is a
correlation between horizontal velocities and the total linear
polarization signals. Especially in the pixels of the FOV where there
are large horizontal velocities we have strong linear polarization
signals. We point out that the mere presence of horizontal velocities
do not imply forward scattering signals, because spatial gradients are
needed to break the axial symmetry. However, a 3D medium is highly
complex and a correlation between $V_h$ and $P_L$ may occur because
thermal, density and velocity field spatial distributions may be
different in regions of the plasma with different levels of dynamical
activity.

\section{The impact of the Hanle effect\label{sec:B}}

In \S4 we already advanced some results regarding the impact of the
Hanle effect on the scattering polarization at the center of the Ca
{\sc i} 4227{\rA} line while comparing the 1.5D and 3D approximations
(see {\figurename} \ref{fig:PL_1.5D_B}). In this section we study in
more detail the impact of the Hanle effect on the fractional linear
polarization $Q/I$, $U/I$, and $P_L$ taking into account the effects
of 3D radiative transfer.

\begin{figure*}
  \centering
  \epsscale{1.04}
  \plotone{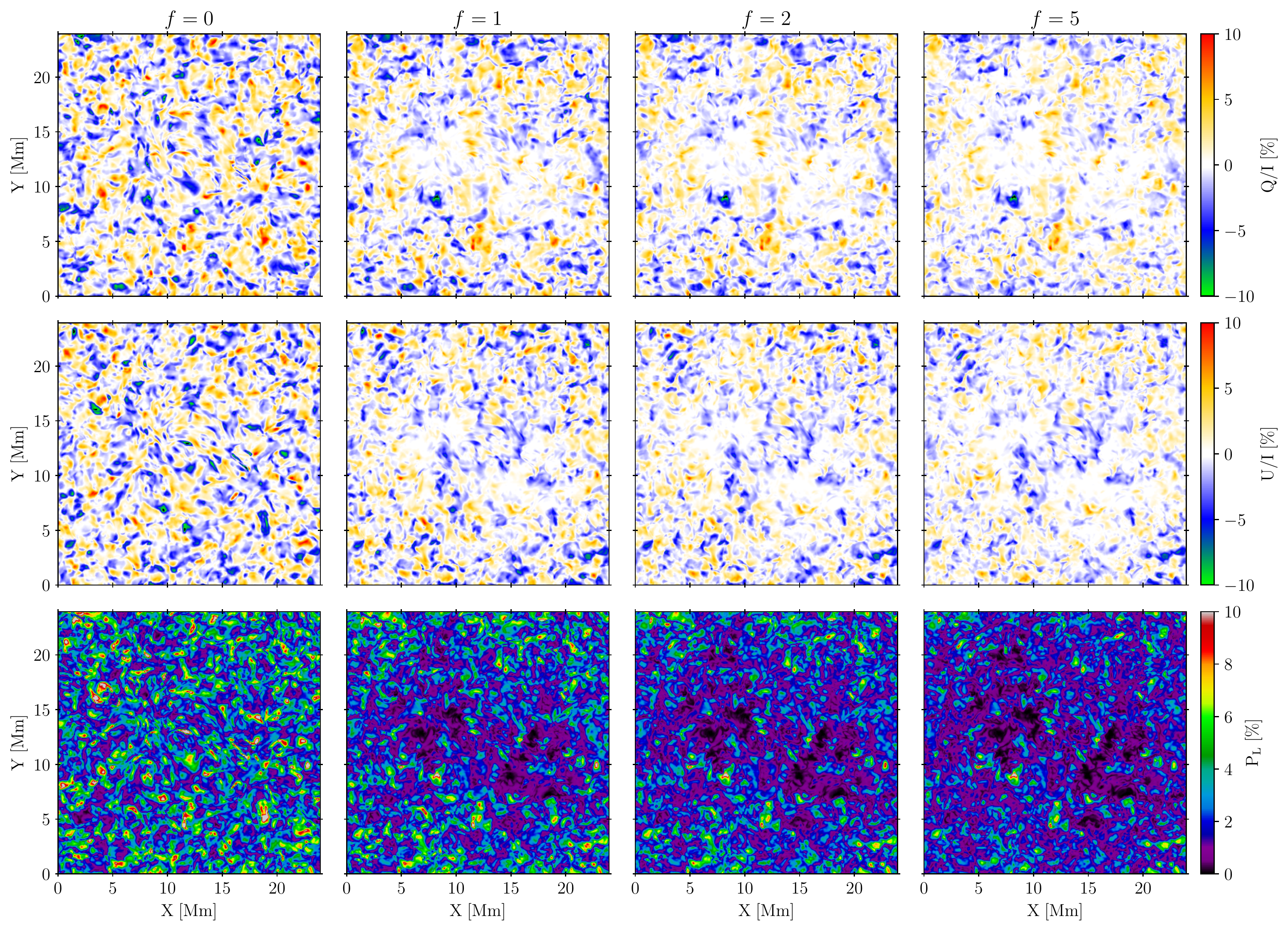}
  \caption{$Q/I$ (top row), $U/I$ (middle row), and $P_L$ (bottom row)
    signals of the emergent radiation in the Ca {\sc i} 4227{\rA}
    line for the full 3D radiative transfer solutions. The different
    columns correspond to $f=0$ (first column), $f=1$ (second column),
    $f=2$ (third column), and $f=5$ (fourth column), with $f$ the
    scaling factor of the model's magnetic field strength. The
    positive reference direction for Stokes $Q$ is along the Y-axis of
    the figures.}
  \label{fig:Hanle-3D}
\end{figure*}

\begin{figure}
  \centering
  \epsscale{1.15}
  \plotone{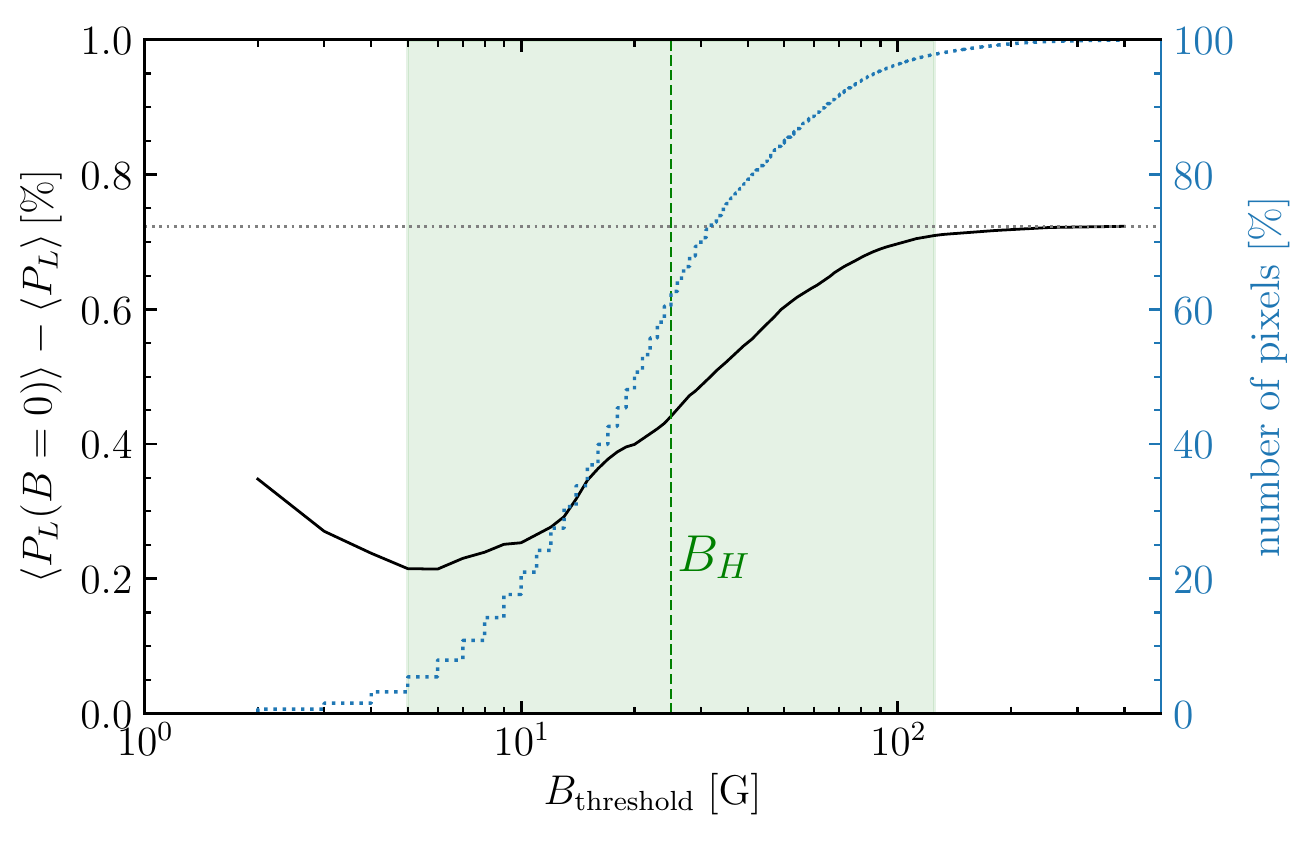}
  \caption{\textbf{Black curve}: average of the indicated $P_L$ signal
    differences considering the pixels where the magnetic field
    strength $B$ at the heigths where $\tau(\lambda_{P_L})=1$ is lower
    than $B_\mathrm{threshold}$. \textbf{Blue curve}: percentage of
    pixels in the FOV that fulfill the condition
    $B<B_\mathrm{threshold}$. The dashed vertical green line labelled
    $B_H$ indicates the critical Hanle field for the Ca {\sc i}
    4227{\rA} line. The shaded region goes from $0.2B_H$ to
    $5B_H$. The dotted horizontal black line shows the difference
    between the two averages considering all the pixels (i.e., the
    full FOV).}
  \label{fig:mean_threshold}
\end{figure}

To this end, we solve the radiative transfer problem in the 3D model
atmosphere, but multiplying the model's magnetic field strength by a
constant scaling factor $f$, namely, 0 (unmagnetized), 1 (true model's
magnetic field), 2, and 5 ({\figurename} \ref{fig:Hanle-3D}). We
clearly appreciate that the Hanle effect tends to strongly depolarize
the zero-field forward scattering signals in the region of the FOV
with the largest magnetic field strengths (i.e., the region between
the two opposite polarities in {\figurename} \ref{fig:Btau}).  The
average of the total linear polarization for the non-magnetic case is
$\langle P_L(B=0) \rangle \approx 3.0\,\%$, while for the magnetic
case ($f=1$) is $\langle P_L \rangle \approx 2.3\,\%$.  However,
outside such magnetic regions the role of the Hanle effect is less
obvious.  Considering only the $P_L$ signals in the pixels where
$B<10$ G at the heights where $\tau(\lambda_{P_L})=1$ in the
\modelB{1}{1}{1}{1} solution we obtain
$\langle P_L(B=0) \rangle \approx 3.1\,\%$ and
$\langle P_L \rangle \approx 2.8\,\%$.  The $P_L$ signals of the
emergent radiation are closer between the non-magnetic and magnetic
solutions in the regions where the magnetic field is relatively weak.
In order to study how the magnetic field strength modifies the linear
polarization, we average the $P_L$ signals considering only the pixels
in the FOV where $B<B_\mathrm{threshold}$ at the heights where
$\tau(\lambda_{P_L})_{\mu=1}=1$ in \modelB{1}{1}{1}{1}.  {\figurename}
\ref{fig:mean_threshold} shows the difference
$\langle P_L(B=0)\rangle - \langle P_L\rangle$ for several magnetic
field thresholds.  The difference starts to increase from
$B_\mathrm{threshold}=5$ G and keep increasing with
$B_\mathrm{threshold}$ up to $B_\mathrm{threshold}=125$ G.  Note that,
via the Hanle effect, the linear polarization is sensitive to magnetic
field strengths between $0.2B_H$ and $5B_H$, which for the Ca {\sc i}
4227 {\rA} line correspond to $5$ G and $125$ G, respectively.
Therefore, we can conclude that most of the forward scattering
polarization signals in the magnetic case are depolarized in regions
where the line is sensitive to the Hanle effect.

On the other hand, the averaged $P_L$ signals for the $f=2$ and $f=5$
cases are $\langle P_L\rangle\approx 1.9\,\%$ and
$\langle P_L\rangle\approx 1.7\,\%$, respectively.  In these cases,
the values are much more similar because for $f\ge 2$ the Ca {\sc i}
4227 {\rA} line is practically in the Hanle saturation regime.

\section{The impact of instrumental effects}

\begin{figure*}
  \centering
  \epsscale{1}
  \plottwo{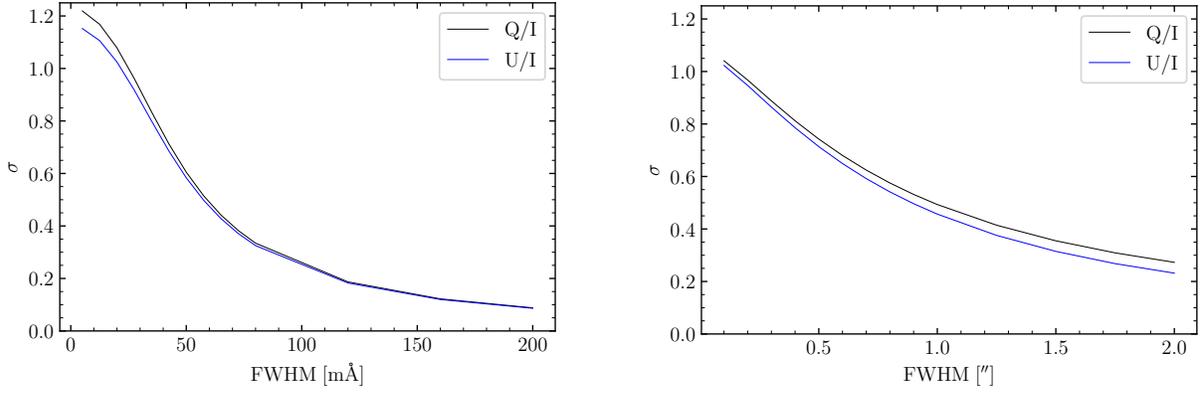}{{Figure_9b}.pdf}
  \caption{\textbf{Left panel}: sensitivity of the standard deviation
    of the $Q/I$ and $U/I$ spatial variations to the spectral
    resolution, assuming a seeing of 0\farcs5. \textbf{Right panel}:
    sensitivity of the standard deviation of the $Q/I$ and $U/I$
    spatial variations to the seeing, assuming a spectral resolution
    of $\Delta\lambda=40\,\rm{m\AA}$.}
  \label{fig:spectral_seeing}
\end{figure*}

One of the many scientific opportunities that the new generation of
solar telescopes will provide is to measure, with unprecedented
resolution, the polarization of chromospheric lines produced by
anisotropic radiation pumping and the Hanle and Zeeman effects. In
this respect, of particular interest are simulations showing what we
may expect to detect using the ViSP instrument at DKIST. To this end,
we have calculated the Stokes profiles of the Ca {\sc i} 4227{\rA}
line at the model's disk-center taking into account the combined
action of scattering processes and the Hanle and Zeeman effects, using
the atomic density matrix elements iteratively computed with PORTA
neglecting the Zeeman splittings.\footnote{This is a suitable
  approximation when the Zeeman splitting is a small fraction of the
  line's Doppler width \citep{1996SoPh..164..117B}, as is indeed the
  case with the used 3D model (see the lower right panel of
  {\figurename} \ref{fig:Btau}).}

The previous sections showed the Stokes signals of the Ca {\sc i}
4227{\rA} line radiation calculated without accounting for any spatial
and spectral degradation. In this section we show the impact that the
finite spatial and spectral resolutions typical of real observations
have on the synthetic Stokes profiles. To this end, we have applied a
code developed by \cite{2018ApJ...863..164D}, which accounts for the
following contributions:

\begin{enumerate}

\item The primary mirror diffraction and the atmospheric seeing,
  modeled applying the long-exposure approximation
  \cite[see][]{1966JOSA...56.1667F}.
  
\item Diffraction in the spectral dimension, modeled by means of a
  convolution with a gaussian having full width at half maximum
  ${\rm FWHM}=\lambda_0/R$, where $R$ is the resolution power of the
  spectrograph and $\lambda_0$ the observed wavelength.

\item Finite width of the spectrograph slit. For each position of the
  FOV covered by the slit, the emergent Stokes profiles for all the
  points across its width are added together.

\item Finite pixel size. The Stokes profiles of the emergent radiation
  for all the points in the surface of the 3D model inside the
  projection of each CCD pixel over the surface are added together.

\item We add some noise by adding random values to $Q$, $U$ and $V$
  following a gaussian distribution with a given standard deviation
  $\sigma_{Q}$, $\sigma_U$, and $\sigma_V$, respectively.

\end{enumerate}

The left panel of {\figurename} \ref{fig:spectral_seeing} shows the
change of the standard deviation of the $Q/I$ and $U/I$ forward
scattering signals for different spectral resolutions at a fixed
0\farcs5 spatial resolution. The standard deviation of the $Q/I$ and
$U/I$ spatial variations is reduced by a factor two for a spectral
resolution of 50 m{\rA}. Clearly, a spectral resolution significantly
better than 100 $\rm{m\AA}$ is needed to detect the spatial
fluctuations of the polarization signals.

Complementary information is provided in the right panel
of {\figurename} \ref{fig:spectral_seeing}, which shows the standard
deviation of the $Q/I$ and $U/I$ disk-center signals for different
spatial resolutions at a fixed 40 $\rm{m\AA}$ spectral resolution. The
panel shows that for a seeing of $1\arcsec$ the standard deviation of
the $Q/I$ and $U/I$ spatial variations is reduced by a factor two. A
spatial resolution better than $2\arcsec$ is essential to detect
spatial variations in the $Q/I$ and $U/I$ disk-center signals.

\begin{figure*}
  \centering
  \epsscale{1.15}
  \plotone{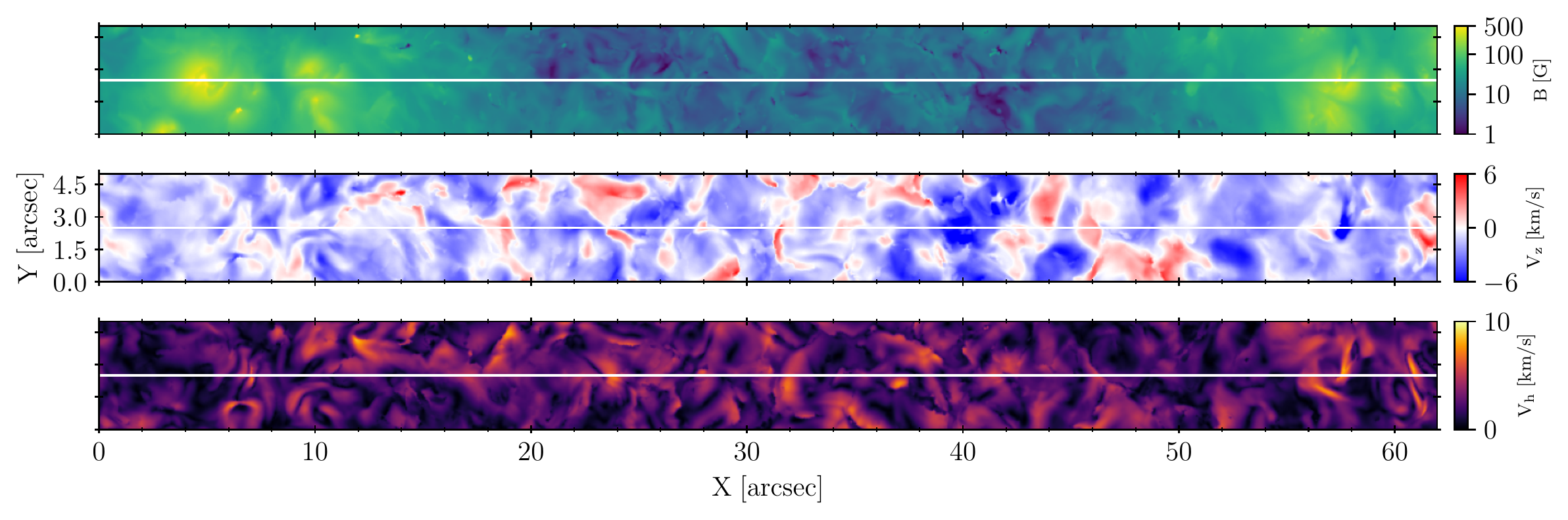}
  \caption{Magnetic field strength (top panel), vertical velocity
    $V_z$ (middle panel), and horizontal velocity $\sqrt{V_x^2+V_y^2}$
    (bottom panel) at the corrugated surface within the 3D model where
    $\tau(\lambda_{P_L})=1$. Each panel shows the region around the
    spectrograph slit indicated with the white line.}
  \label{fig:par_slits}
\end{figure*}

The white line in {\figurename} \ref{fig:par_slits} visualizes the
spectrograph's slit covering $62\arcsec$ on the surface of the 3D
model.  As seen in the figure, we have oriented the slit so as to
cross a few patches of magnetic flux concentrations having field
strengths exceeding $100\,\rm{G}$ (see the upper panel in
{\figurename} \ref{fig:par_slits}).  We can also find significant
(several km/s) macroscopic velocities, both vertical and horizontal,
along the spatial direction of the spectrograph's slit at the
chromospheric heights where the optical depth at the wavelength
$\lambda_{P_L}$ is unity (see middle and bottom panels in
{\figurename} \ref{fig:par_slits}).

\begin{figure*}
  \centering
  \gridline{\fig{{Figure_11_1.0e-04_5s}.pdf}{0.9\textwidth}{(a)
      Noise PDF $\sigma=10^{-4}$, $5\,\rm{s}$ exposure time}}
  \gridline{\fig{{Figure_11_1.0e-05_493s}.pdf}{0.9\textwidth}{(b)
      Noise PDF $\sigma=10^{-5}$, $490\,\rm{s}$ exposure time}}
  \caption{Simulation of slit spectropolarimetric observations with
    ViSP at DKIST. \textbf{Top}: for an exposure time of $5\,\rm{s}$,
    which implies a noise PDF of $\sigma=10^{-4}$. \textbf{Bottom}:
    for an exposure time of $490\,\rm{s}$, which implies a noise PDF
    of $\sigma=10^{-5}$. In each pair of panels (for $Q/I$, $U/I$ or
    $V/I$) the right one shows the non-magnetic case, while the left
    one includes the impact of the model's magnetic field through the
    Hanle and Zeeman effects. The reference direction for $Q>0$ is
    along the spatial direction of the slit. The color tables are
    saturated at $\pm 1\%$.}
  \label{fig:visp}
\end{figure*}

The ViSP instrument offers several configurations. We have selected
the second arm with a spatial sampling of $0\farcs0242$ and the
$0\farcs2142$ wide slit, with a spectral sampling of
$8.4\,\rm{m\AA}$. With this set-up the resolution power is
$R=70\,000$. We characterize the noise in the polarization quantities
with the standard deviation ($\sigma_{QUV}$) of a gaussian probability
distribution function (PDF) given in units of the continuum
intensity. The width of the noise distribution depends on the exposure
time; thus, we have chosen the following two: $t=5\,\rm{s}$
($\sigma_{QUV}=4.5\cdot10^{-4}$) and $t=490\,\rm{s}$
($\sigma=4.5\cdot10^{-5}$).  \footnote{All the instrument-dependent
  parameters are extracted from the \textit{Instrument Performance
    Calculator} of the ViSP instrument:
  \url{https://nso.edu/telescopes/dkist/instruments/visp/}}

The spatial resolution is typically limited by the atmospheric seeing,
which we assume to be $0\farcs5$. Since the pixel size is below the
spatial resolution, we bin 10 pixels along the slit increasing the
signal to noise ratio (SNR). Consequently, our simulated observations
have a spatial and spectral resolution of $0\farcs5$ and
$60\,\rm{m\AA}$, respectively, and the noise PDF turns out to be
$\sigma_{QUV}=10^{-4}$ for $t=5\,\rm{s}$ and $\sigma_{QUV}=10^{-5}$
for $t=490\,\rm{s}$.

{\figurename} \ref{fig:visp} shows the $Q/I$, $U/I$ and $V/I$
profiles, at each point along the slit, corresponding to the
above-mentioned ViSP instrumental setup and for $t=5\,\rm{s}$ (top
panel) and $t=490\,\rm{s}$ (bottom panel) exposure times. Each panel
in the figure shows three pairs of plots, one for each fractional
Stokes parameter, and with each pair showing the cases where we take
into account (left) or neglect (right) the model's magnetic field. As
expected, the $V/I$ Zeeman signals are very significant where we have
the largest longitudinal magnetic field strengths, showing signals
above 1\% (see the left side of the right column in {\figurename}
\ref{fig:visp}). In such regions the Hanle effect tends to depolarize
the non-magnetic linear polarization signals (see left and middle
columns in {\figurename} \ref{fig:visp}). Outside such ``strong
field'' patches the model's magnetic field is too weak so as to
produce a very clear impact on the $Q/I$ and $U/I$ signals, neither
through the Hanle nor the Zeeman effects. We find that the linear
scattering polarization signals are dominated by the breaking of the
axial symmetry produced by the spatial gradients of the horizontal
components of the model's macroscopic velocities.  Moreover, in the
pixels where the magnetic field strength is around the critical Hanle
field ($B_{\rm H}={25\,\mathrm{G}}$) the Hanle effect significantly
modifies the scattering polarization signals.  For example, at
position $42\arcsec$ along the slit we see a non-magnetic positive
$Q/I$ signal of about 1\% (left side in left panel). This is
considerably reduced by a $\sim 75$ G magnetic field (right side in
left panel), which is not sufficient to generate significant circular
polarization via the Zeeman effect.

With its 4 meter mirror the collecting surface of DKIST is
significantly larger than in the previous generation of solar
telescopes, such as the 1.5 meter mirror of the GREGOR telescope
operating at the Observatorio del Teide (Canary Islands, Spain). The
Z\"urich Imaging Polarimeter (ZIMPOL) attached to GREGOR could be used
for doing spectropolarimetric observations of the Ca {\sc i} 4227
\r{A}\ line with a spatial and spectral sampling of $0\farcs33$ and
$7.9\,\rm{m\AA}$, respectively, giving a spectral resolution power of
$200\,000$ with the $0\farcs3$ slit. We can expect a noise PDF of
about $\sigma_{QUV}=10^{-4}$ for $15\,\rm{min}$ exposure
time. Reaching $\sigma_{QUV}=4\cdot10^{-5}$ with ZIMPOL at GREGOR
would require 2 hours of exposure time. Clearly, the large collecting
area of DKIST is a big advantage, although it remains to be seen if
such a high polarimetric sensitivity can be reached without the very
fast polarization modulation that is possible with the ZIMPOL
instrument.

\begin{figure*}
  \centering
  \gridline{\fig{{Figure_12_1.0e-03_7s}.pdf}{\textwidth}{(a)
      Noise PDF $\sigma=10^{-3}$, $7\,\rm{s}$ exposure time.}}
  \gridline{\fig{{Figure_12_2.0e-04_163s}.pdf}{\textwidth}{(b)
      Noise PDF $\sigma=2\cdot10^{-4}$, $163\,\rm{s}$ exposure time.}}
  \caption{Simulation of slit spectropolarimetric observations with
    ViSP at the diffraction limit of the DKIST telescope. The spatial
    resolution is $0\farcs0398$ and the spectral resolution
    $14.6\,\rm{m\AA}$. The reference direction for $Q>0$ is along the
    spatial direction of the slit.}
  \label{fig:visp_diff}
\end{figure*}

When observing at the diffraction limit of a telescope the SNR is
independent of its aperture \citep[e.g.,][]{article}. It is of
interest to consider now the case of spectropolarimetric observations
with ViSP at the spatial resolution limit of DKIST
($0\farcs026$). Obviously, matching this resolution is only possible
if the adaptive optics (AO) system can correct at this precision
during the whole integration time. The minimum spatial sampling can be
achieved with arm 3, $0\farcs0199$ covering $51\,\rm{arcsec}$ along
the slit, giving a resolution of $0\farcs0398$. With the narrowest
slit ($0\farcs0284$), we get a spectral sampling of $7.3\,\rm{m\AA}$
and a spectral resolution of $\Delta \lambda=14.6\,\rm{m\AA}$. The top
panels of {\figurename} \ref{fig:visp_diff} show the simulated
observation corresponding to this configuration, with an exposure time
of $7$ second resulting in a noise PDF with
$\sigma_{QUV}=10^{-3}$. The bottom panels show the case for an
exposure time of $2.7\,\mathrm{min}$ resulting in a noise PDF with
$\sigma_{QUV}=2\cdot10^{-4}$. Clearly, achieving a high enough SNR is
important for a clear detection and quantification of the weakest
polarization signals, along with their spatial variability. We point
out that the temporal evolution of the chromospheric plasma has not
been considered here, because full 3D radiative transfer calculations
are computationally costly; the signals and spatial variations
shown in the figure would decrease as the exposure time of the
simulated observation increases.

\section{Summary and conclusions\label{sec:C}}

We have carried out a detailed 3D radiative transfer investigation of
the linear polarization signals produced by forward scattering
processes in the core of the Ca {\sc i} 4227 \r{A}\ line, sensitive to
chromospheric magnetic fields via the Hanle effect. To this end, we
have applied PORTA, a 3D radiative transfer code for modeling the
intensity and polarization of spectral lines with massively parallel
computers \citep{2013A&A...557A.143S}. The solar model atmosphere used
is a 3D snapshot model resulting from a radiation magneto-hydrodynamic
simulation by \cite{2016A&A...585A...4C}, which is representative of
an enhanced-network region.

We show that the line's Doppler shifts resulting from the spatial
gradients of the horizontal components of the plasma macroscopic
velocities break the axial symmetry of the radiation field that
illuminates each point within the model chromosphere, to the extent
that they produce sizable forward scattering polarization in the core
of the Ca {\sc i} 4227 \r{A}\ line without the need of any magnetic
field (see {\figurename} \ref{fig:QUP_0B0V_0B0Vh_0B0Vz}). The spatial
gradients of the vertical components of the macroscopic velocities
must also be taken into account because the ensuing line's Doppler
shifts modify the radiation field anisotropy and the scattering
polarization signals. Interestingly, the correlation between $V_h$
and $P_L$ seen in {\figurename} \ref{fig:histV} suggests that regions
of the solar chromosphere showing large horizontal velocities may also
show strong forward-scattering polarization signals in the Ca {\sc i}
4227 \r{A}\ line.

The Hanle effect caused by the model's magnetic field tends to
depolarize the $Q/I$ and $U/I$ disk-center signals at the locations of
the field of view where $5$ G $<B<125$ G in the lower chromosphere
(see {\figurename}s \ref{fig:Hanle-3D} and \ref{fig:mean_threshold}).
However, in the regions where the magnetic field is weaker, the Hanle
effect just modifies the emergent linear polarization without a
dominant tendency. The Ca {\sc i} 4227 {\rA} line is in the Hanle
saturation regime for $B>125$ G. In such regime, the linear
polarization is no longer sensitive to the magnetic field strength,
but only to its orientation.

The 1.5D approximation, when applied retaining only the vertical
components of the plasma macroscopic velocities, is unsuitable for
modeling the forward scattering polarization of this line, because it
leads to the wrong conclusion that detection of polarization implies
the presence of inclined magnetic fields. We have accounted for the
symmetry breaking produced by the horizontal components of the plasma
macroscopic velocities in the 1.5D approximation, which provides
results in qualitative agreement with the full 3D calculations (see
{\figurename} \ref{fig:IQU_1.5D0Vh_1.5D_3D}). Full 3D radiative
transfer is however needed for an accurate quantification of the
scattering polarization signals.

Finally, our simulations of the scattering polarization signals of the
Ca {\sc i} 4227 \r{A}\ line taking into account the degradation of
instrumental effects show that a spectral resolution and a spatial
resolution better than 100 m{\rA} and 2$\arcsec$ are essential to
detect spatial variations of the disk-center polarization signals (see
{\figurename} \ref{fig:spectral_seeing}). The Visible
Spectro-Polarimeter (ViSP) at the DKIST achieves such requirements
with a very high SNR in few seconds. We have simulated
spectropolarimetric observations with the ViSP with a spatial
resolution limited by seeing (see {\figurename}
\ref{fig:visp}). Considering an integration time of 5 seconds, we are
able to detect the spatial variations of the strongest scattering
polarization signals. As we increase the exposure time, the SNR is
larger and we can then appreciate much better the rich spatial
variability on the forward-scattering polarization signals. To observe
at the spatial resolution limit of the telescope (see {\figurename}
\ref{fig:visp_diff}) we would need at least 2 minutes of exposure time
to detect the spatial variations with sufficient SNR. Finally, it is
important to note that we have not considered the temporal evolution
of the chromospheric plasma; therefore, as we increase the exposure
time the signals would decrease due to signal cancellations and the
spatial variations would be more difficult to detect.

Clearly, a reliable modeling of the linear polarization produced by
scattering processes in the Ca {\sc i} 4227 \r{A}\ chromospheric line
requires taking into account all the causes that may break the axial
symmetry of the pumping radiation field, both magnetic and
non-magnetic. In this paper we have taken all such symmetry breaking
causes into account showing how the line's scattering polarization is
sensitive to the magnetic, thermal and dynamic structure of the lower
solar atmosphere. The radiative transfer tools (see the public
repository of PORTA at \url{https://gitlab.com/polmag/PORTA}) needed
for this type of modeling are at the disposal of the astrophysical
community for validating or refuting the 3D models of the solar
chromosphere, by means of contrasting the calculated Stokes profiles
with the unprecedented spectropolarimetric observations that the new
generation of solar telescopes will hopefully provide.

\acknowledgements J.J.B. acknowledges financial support from the
Spanish Ministry of Economy and Competitiveness (MINECO) under the
2015 Severo Ochoa Programme MINECO SEV--2015--0548. J.T.B. and
T.P.A. acknowledge the funding received from the European Research
Council (ERC) under the European Union's Horizon 2020 research and
innovation programme (ERC Advanced Grant agreement \mbox{No.~742265}),
as well as through the projects PGC2018-095832-B-I00 and
PGC2018-102108-B-I00 of the Spanish Ministry of Science, Innovation
and Universities.  J.\v{S}. acknowledges financial support through
grant \mbox{19-20632S} of the Czech Grant Foundation (GA\v{C}R) and
project \mbox{RVO:67985815} of the Astronomical Institute of the Czech
Academy of Sciences. Likewise, J.T.B. and J.\v{S}. acknowledge the
support received from the Swiss National Science Foundation through
grant CRSII5-180238. The 3D radiative transfer simulations were
carried out with the MareNostrum supercomputer of the Barcelona
Supercomputing Centre (National Supercomputing Centre, Barcelona,
Spain), and we gratefully acknowledge the technical expertise and
assistance provided by the Spanish Supercomputing Network, as well as
the additional computer resources used, namely the La Palma
Supercomputer located at the Instituto de Astrof\'isica de Canarias.

\bibliography{mybibtex}{}
\bibliographystyle{aasjournal}

\end{document}